\documentclass[%
 reprint,
 amsmath,amssymb,
 aps,
floatfix,
]{revtex4-2}

\usepackage{graphicx}
\usepackage{dcolumn}
\usepackage{bm}
\usepackage{physics}
\usepackage{amsmath}
\usepackage{natbib}
\usepackage{amssymb}
\usepackage{romannum}
\usepackage{array}
\usepackage{xcolor}
\usepackage{soul}
\usepackage[normalem]{ulem}

\setstcolor{red}
\usepackage{lineno}

\begin{document}

\preprint{APS/123-QED}

\title{Excitation-pulse intensity mediated control of coherent nonlinear optical response of a V-type system}

\author{Rishabh Tripathi}
\affiliation{Department of Physics, Indian Institute of Science Education and Research Bhopal, Bhopal 462066, India}
\author{Krishna K. Maurya}
\affiliation{Department of Physics, Indian Institute of Science Education and Research Bhopal, Bhopal 462066, India}
\author{Rohan Singh}
\email{rohan@iiserb.ac.in}
\affiliation{Department of Physics, Indian Institute of Science Education and Research Bhopal, Bhopal 462066, India}

\email{\authormark{}rohan@iiserb.ac.in}
\date{\today}

\begin{abstract}
V-type three-level systems, where two excited states share a common ground state, serve as fundamental models for exploring coherent light-matter interactions in a range of quantum systems, from atomic gases to semiconductor nanostructures. 
In this work, we investigate the coherent evolution of such a system under strong femtosecond-pulse excitation by numerically solving the optical Bloch equations.
Our analysis shows that the coherent evolution of a three-level system critically depends on the product of the excitation-pulse duration and energy separation between the excited states. 
Building on this understanding, we extend our analysis to simulate two-dimensional coherent spectra in a high-intensity regime.
We demonstrate a control over the coherent pathway contributions to the nonlinear optical response of a V-type system by varying the intensity of the excitation pulses.
This control is manifested through the ability to selectively turn individual spectral features on or off in the 2D spectra, each corresponding to distinct quantum pathways.
Furthermore, the pulse intensities are varied to precisely adjust the phase of these peaks.
Our approach provides a simple and robust framework for achieving control of coherent response of multilevel systems.

\end{abstract}

\maketitle



\section{Introduction}
The ability to control the coherent optical response of quantum systems is fundamental for applications in quantum information science \cite{Chen2001,Biolatti2000,Sussman2005,Monmayrant2006}. 
At its core, this method involves steering quantum states toward desired outcomes through their interaction with light. 
Optical coherent control has been demonstrated in a variety of systems ranging from atoms and molecules to semiconductors, supporting technological developments in several fields.
For instance, it enables precise control over chemical reactions \cite{Assion1998}, facilitates quantum computing through state manipulation \cite{Weinacht1999a}, allows for rapid manipulation of electronic states in nanostructures \cite{Kamada2001,ViasnoffSchwoob2005,Borri2002,Beirne2006,Htoon2002}, and switching between and modifying specific coherent nonlinear signals \cite{Fras2016, Lim2011}, which is pivotal for advancing optoelectronic devices.

A key mechanism for demonstrating coherent control in quantum systems is through Rabi oscillations. 
This phenomenon has been extensively studied in two-level systems (TLSs) \cite{Allen1975} under both continuous-wave (CW) \cite{Flagg2009} and pulsed-laser excitation \cite{Stievater2001,Zrenner2002} where periodic population transfer occurs between quantum states under a coherent optical field.
If a TLS with transition frequency $\omega$ is excited by a CW laser with electric field amplitude $E_0$ and frequency $\omega_{0}$, the excitation probability oscillates at a frequency given by the generalized Rabi frequency \cite{Cundiff1994}
\begin{equation}
    \Omega_{eff} = \sqrt{{\Delta_{\omega}}^2 + \left(\frac{\mu {E_0}}{\hbar}\right)^2},
\end{equation}
where $\Delta_{\omega} = \omega - \omega_{0}$ is the frequency detuning and $\mu$ is the transition dipole moment. 
In contrast, pulsed lasers introduce a time-dependent electric field envelope $E(t)$, often modeled as a Gaussian, where the instantaneous Rabi frequency can be defined. 
In such cases, Rabi oscillations are more suitably described in terms of the pulse area
\begin{equation}
    \Theta = \frac{1}{\hbar}\int_{-\infty}^{\infty} \mu E(t) \,dt, 
    \label{eq:pa}
\end{equation}
which results in an excitation probability of excited state $P_e \propto \sin^2(\Theta/2)$ \cite{Wigger2017}.
While Rabi oscillations are well understood in TLSs, multilevel quantum systems often involve multiple transitions that cannot be simultaneously excited by a single CW laser due to its narrow spectral bandwidth.
Ultrafast pulses, with their broad bandwidth and high peak intensity, enable simultaneous excitation of multiple states and can induce Rabi-oscillations-type coherent evolution, making pulsed lasers a powerful tool for coherent control beyond a TLS.

\begin{figure}
    \centering
    \includegraphics{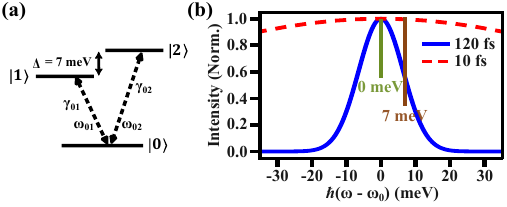}
    \caption{(a) The energy levels in a V-type system, illustrating the ground state $\ket{0}$, first excited state $\ket{1}$, and second excited state $\ket{2}$, with their respective coherence decay rates $(\gamma_{01},\: \gamma_{02})$ and transition frequencies $(\omega_{01},\: \omega_{02})$.
    $\Delta$ is the energy difference between the two excited states.
    The dipole allowed transitions are indicated by the dashed arrows.
    (b) The spectra of laser pulses with duration 10 fs and 120 fs are shown with the positions of the excited states at 0 meV and 7 meV within the laser spectrum.}
    \label{fig:pulse}
\end{figure}

A V-type three-level system \cite{Wang2001,Albrecht1996,Ficek2004} is an extension of a TLS comprising two excited states coupled to a common ground state, as shown in Fig. \ref{fig:pulse}(a). 
It provides a valuable model for exploring exotic quantum dynamics arising from the interference of multiple quantum pathways \cite{Lee1997}. 
Nevertheless, simultaneous excitation of two nondegenerate transitions can lead to intricate population dynamics due to coupled-state interactions, making coherent control more challenging.
Previous studies have demonstrated coherent control in V-type systems by polarization selection \cite{Wang2005,Farrell2022}, using chirped pulses \cite{Lim2011,Wollenhaupt2005} or other pulse-shaping techniques \cite{Tollerud2014, Marroux2016, Meshulach1998,Grisard2022}.
However, chirped pulses can distort spectral lines \cite{Tekavec2010, Binz2020}, and polarization-based methods are often ineffective when excited-state transitions lack polarization sensitivity.
Furthermore, pulse-shaping techniques can be experimentally complex to implement.
These limitations underscore the demand for a simpler and more universally applicable approach to control the coherent evolution of multilevel systems.

Multiple pathways may contribute to the coherent nonlinear response of a system.
Nonlinear spectroscopy techniques based on wave-mixing are used isolate specific signal contributions for a better understanding of the system.
Additionally, wave-mixing experiments have also been used to demonstrate coherent control via pathway selection \cite{Fras2016} or enhancement \cite{Lim2011}.
Two-dimensional coherent spectroscopy (2DCS) is an example of a wave-mixing technique where the system's response is spread across two frequency axes \cite{Jonas2003,Li2023}; specific quantum pathways appear as distinct peaks in a 2D spectrum.
A control over these peaks by isolating or modifying these spectral features is a clear demonstration of the control of the coherent response of the system.
Such coherent control in 2DCS experiments has previously provided access to structural information \cite{zanni2001}, reaction selectivity \cite{Lim2011}, and coherence transfer mechanisms \cite{Marroux2016}.
Previous studies of V-type systems have used high-intensity excitation pulses to modify the peaks in 2D spectra due to higher-order nonlinear contributions \cite{Binz2020, Chen2017}.
However, the potential of exploiting this intensity-dependent modification to achieve coherent pathway control in 2DCS has not been explored.
Inspired from these studies, we propose a straightforward approach to control the coherent response of multilevel systems by using intense, finite-duration excitation pulses in 2DCS.
Our study highlights the ability to precisely manipulate both the amplitude and phase of the peaks in 2D spectra by employing intense laser fields, demonstrating a high degree of control over the quantum pathways that define the system's evolution upon interaction with light.
This method reveals a new mechanism to control the coherent dynamics by exploiting intensity-dependent nonlinear interactions.

The paper is organized as follows: Sec. \ref{V-type} provides details of a model V-level system that we have considered.
Section \ref{LMI_1} investigates the coherent evolution of the V-type system under single-pulse excitation. 
For a spectral width comparable to the energy gap between excited states, high-intensity pulses induce a rich and nonintuitive coherent dynamics.
This effect is absent in the impulsive limit. 
These complex dynamics are then leveraged to demonstrate coherent control through 2DCS simulations, which are detailed in Secs. \ref{Per_vs_NonPer} \& \ref{Results}. 
The paper concludes with a discussion of the results from our 2DCS simulations.

\section{V-level system}
\label{V-type}

In our study, we model a V-type three-level system, as illustrated in Fig. \ref{fig:pulse}(a). 
This system features two excited states, labeled $\ket{1}$ and $\ket{2}$, separated by an energy difference $\Delta=7$ meV, which corresponds to a frequency difference $\Delta_{\nu}=1.7$ THz. 
Both excited states have a common ground state $\ket{0}$.
Relevant parameters such as $\Delta$, relaxation rates, and transition dipole moments are similar to those for the D$_1$ and D$_2$ transititons in potassium \cite{Li2013}.
These parameters are detailed in Table \ref{tab:Param}. 

\begin{table}
    \caption{\label{tab:Param} Parameters used for simulation.}
    \begin{ruledtabular}
    \begin{tabular}{lc}
        Dephasing rate ($\gamma_{01}, \gamma_{02}$) & 0.193 meV \\ 
        Dephasing rate ($\gamma_{12}$) & 0.386 meV \\ 
        Population decay rate ($\gamma_{1}, \gamma_{2}$) & 0.386 meV \\ 
        Energy separation ($\Delta$) & 7 meV \\ 
        Transition dipole moment ratio $\left(\mu_{02}/\mu_{01}\right)$ & 1.4 \\  
    \end{tabular}
    \end{ruledtabular}
\end{table}

The following sections explore how intense, finite-duration pulses affect the coherent evolution of our model system.
We start by exciting this system with a single, intense pulse. 

\section{Single-pulse excitation}
\label{LMI_1}
A single optical pulse can simultaneously excite both the excited states of a V-level system if the spectrum is broad enough to cover both transition frequencies.
Consequently, we can induce Rabi-oscillations-type population evolution of the excited states using a sufficiently intense laser pulse.
Although direct transitions between the excited states are forbidden by dipole selection rules, their coupling through the common ground state affects the system's coherent response. 
For the sake of brevity, we will refer to these phenomena as Rabi oscillations of the V-level system in the rest of the paper.

The light–matter interaction for a V-type system is described by a set of equations similar to the optical Bloch equations (OBEs) for a TLS. 
The extended OBEs for the V-level system include the coherent superposition of the excited states, allowing us to capture the coupling between them accurately; these equations are detailed in Appendix \ref{app:OBEs_V}. 
These OBEs can be solved analytically for $\delta$-function pulse excitation to obtain a result similar to Eq. \eqref{eq:pa} with $\mu$ replaced by $\mu_{eff}=\sqrt{\mu_{01}^2 + \mu_{02}^2}$ \cite{Wang2005}; thus, $\mu_{eff} = 1.72 \mu_{01}$ for the chosen system parameters.
Consequently, by defining $\Theta$ with respect to the first transition, we obtain a Rabi period of $1.16\pi$.

\begin{figure}
\includegraphics{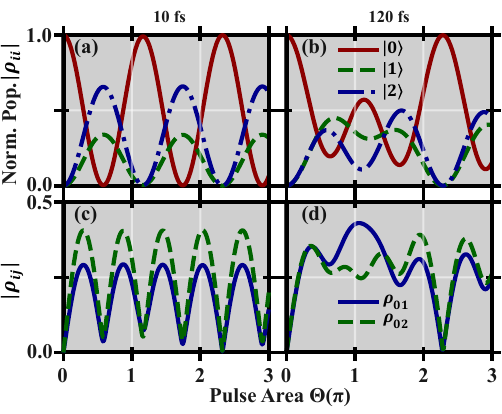}
\caption{Single-pulse excitation of a V-level system. Population of all three states ($\ket{0}$, $\ket{1}$, and $\ket{2}$) as a function of pulse area after excitation with a (a) 10-fs and (b) 120-fs Gaussian pulse.
Polarizations ($\abs{\rho_{01}}$, and $\abs{\rho_{02}}$) after excitation with the (c) 10-fs and (d) 120-fs pulse.
Note that pulse areas are defined relative to the transition dipole moment $\mu_{01}$ of the $\ket{0} \leftrightarrow \ket{1}$ transition.
}
\label{fig:Norm_pop}
\end{figure}

These OBEs cannot be solved analytically for finite-duration Gaussian pulses.
We solve them numerically and monitor the evolution of the system’s density matrix as a function of pulse area.
We assume that the laser pulse is resonant with the transition to the first excited state, while the second excited state is detuned by $\Delta$. 
We focus on two representative cases: excitation by a very short Gaussian pulse with FWHM of 10 fs, which approximates a $\delta$-function pulse, and a comparatively longer 120 fs Gaussian pulse.
Figure \ref{fig:Norm_pop} illustrates both the scenarios.
In the case of the 10 fs pulse, the populations of all three states exhibit clear, sinusoidal oscillations with respect to the pulse area, as shown in Fig. \ref{fig:Norm_pop}(a).
This behavior shows that the two transitions evolve in sync with each other with the period of oscillations equal to the analytical value of the Rabi period obtained for $\delta$-function excitation.
The coherences $|\rho_{01}|$ and $|\rho_{02}|$ also demonstrate sinusoidal Rabi oscillations, as shown in Fig. \ref{fig:Norm_pop}(c).
The period appears half of that for the populations since the phase of coherences $\rho_{01}$ and $\rho_{02}$ flips during a single Rabi oscillation.

In contrast, the Rabi oscillations become significantly more complex when the system is excited with a longer 120 fs pulse, as illustrated in  Figs. \ref{fig:Norm_pop}(b) \& \ref{fig:Norm_pop}(d).
The observed pattern is characterized by multiple local maxima and minima across various pulse areas.
We note that, because of the complex coherent dynamics, the pulse area values are not very useful in predicting the final state of the system; instead, they primarily quantify the intensity of the excitation pulses.

An obvious consequence of a longer excitation pulse is that the spectral intensity at the frequency of the detuned transition is smaller compared to the spectral peak of the pulse, as shown in Fig. \ref{fig:pulse}(b).
Since the product $\mu E$ shows up in the OBEs, the reduced electric-field amplitude would affect the system's response.
We can mathematically include the effect of finite bandwidth even in $\delta$-function calculations; the reduced $E$ at the detuned transition is accounted for by assigning a smaller value of $\mu$ to that transition ($\mu_{02}$ here), such that the product $\mu E$ remains unchanged.
While this mathematical manipulation can account for pulse-intensity variation across the transitions, it cannot account for the complex dynamics that we observe for the finite-duration excitation pulse.
For instance, we observe the complex dynamics although $\mu_{01}E(\omega_{01}) \approx \mu_{02}E(\omega_{02})$ for the system parameters used here.

Another consequence of the longer excitation pulse is that the two transitions pick up different phases during the excitation process.
We estimate this relative phase by the product $\tau\Delta$, where $\tau$ is the duration of the excitation pulse.
For a 10 fs pulse, $\tau \Delta = 0.017$, which is sufficiently small, causing minimal phase effects and allowing the evolution to remain sinusoidal.
However, the product $\tau \Delta$ increases to 0.2 for the 120 fs pulse, which induces a substantial relative phase between the excitations of the two transitions and significantly affects the system's evolution.
Thus, we attribute this transition from simple sinusoidal to complex, nonsinusoidal behavior to the relative phase acquired by the two excited states during the excitation process.



We further validate the above inference by changing $\Delta$ while keeping $\tau = 120$ fs such that $\tau \Delta = 0.017$.
The Rabi oscillations observed under these conditions (data not shown) are identical to those in Figs. \ref{fig:Norm_pop}(a) and \ref{fig:Norm_pop}(c), highlighting the role of the value of $\tau \Delta$ in determining the coherent evolution provided the other system parameters are unchanged.
Although the phase and amplitude effects go hand in hand, the ad hoc separation of their effects provides an intuitive reasoning behind the observed complex behavior.


\begin{figure}
\includegraphics{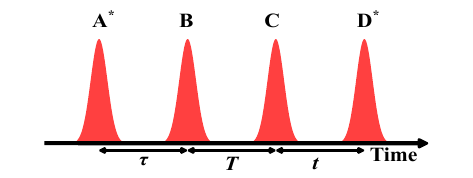}
\caption{Pulse scheme used in simulations. Delays between pulses $\text{A}^*$, B, C, and $\text{D}^*$  are denoted as $\tau$, $T$, and $t$, respectively. Time zero is defined by arrival time of first pulse.}
\label{fig:Pulse_sequence}
\end{figure}

\begin{figure*}
\includegraphics{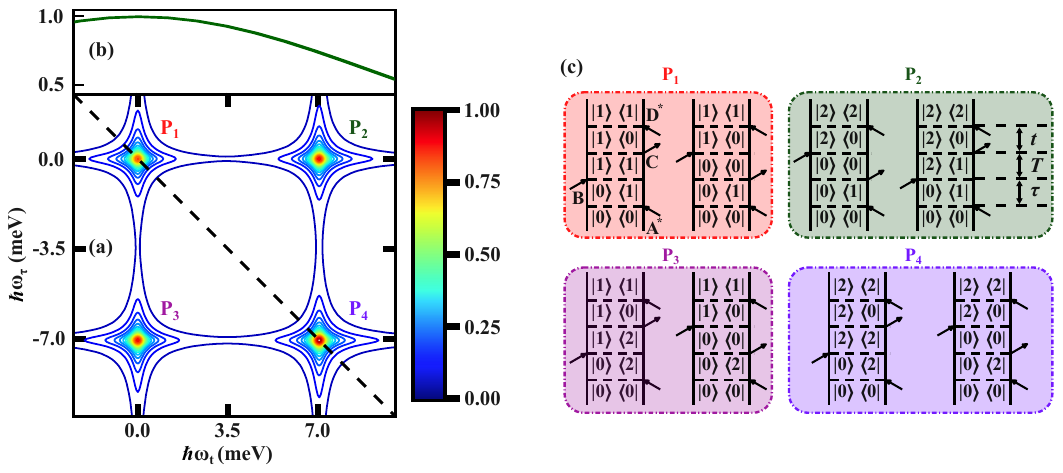}
\caption{ (a) Simulated rephasing 2D spectra for V-level system in the perturbative regime. (b) Normalized excitation-pulse spectrum. (c) Feynman diagrams labeled for each peak in the 2D spectra within the perturbative regime.}
\label{fig:2D_spectra}
\end{figure*}

A key feature of the coherent evolution for nonnegligible values of $\tau \Delta$ is that we can switch between the higher population state and/or the coherence after optical excitation by simply varying the pulse area.
This behavior opens up opportunities for controlled manipulation of quantum pathways; we utilize it to demonstrate coherent control in 2DCS.

\section{2D Simulation: Perturbative vs. non-perturbative regime}
\label{Per_vs_NonPer}
2DCS \cite{Jonas2003,Li2023}, which is an extension of the four-wave mixing (FWM) technique, is commonly employed to investigate coherence dynamics in quantum systems.
The nonlinear response of a material is measured as a function of two time delays between a sequence of phase-coherent ultrafast laser pulses.
A subsequent Fourier transform with respect to both time variables generates a two-dimensional frequency domain spectrum, clearly resolving distinct quantum pathways and their associated interactions.

In our study, we simulate the rephasing 2D spectra of our model V-type system which was described in Sec. \ref{V-type} using a collinear sequence of four Gaussian pulses with $\tau =$ 120 fs, as illustrated in Fig. \ref{fig:Pulse_sequence}. 
Here, we solve the OBEs for a V-level system under a sequence of excitation pulses and employ a phase-cycling approach. 
Specifically, we implement a $3\times3\times3\times1$ phase cycling scheme \cite{Tan2008}, which isolates the desired rephasing signal defined by the phase-matching condition $\phi_s= - \phi_1 + \phi_2 + \phi_3 - \phi_4$.
The rephasing signal is calculated as a function of delays $\tau$ and $t$, which is then Fourier transformed to obtain the rephasing frequency-domain 2D spectrum.
A nonzero value of $T = 590$ fs is used to avoid temporal overlap between pulses B and C.
This value is reciprocal of the detuning $\Delta$ and the results should be identical to those for $T=0$, as discussed in Appendix \ref{app:Role_T}.
These calculations are nonperturbative and simulate the exact nonlinear response and are consistent with 2DCS experiments performed with high-intensity excitation pulses \cite{Tripathi2025}.

Initially, we investigate a simplified scenario to examine the effect of the pulse area of the first excitation pulse $\Theta_1$ on 2D spectra.
In this case, simulations were performed by varying $\Theta_1$ while keeping the subsequent pulses fixed at a low intensity of 0.1$\pi$, which is expected to be within the third-order perturbative regime. 
At low $\Theta_1$ values, the spectral features of the rephasing 2D spectra can be accurately described using double-sided Feynman diagrams (DSFDs). 
Figure \ref{fig:2D_spectra}(a) shows a representative simulated 2D spectrum at low intensity, featuring two diagonal peaks ($P_{1}$ and $P_{4}$) and two cross-peaks ($P_{2}$ and $P_{3}$).
Each peak in Fig. \ref{fig:2D_spectra}(a) is associated with two corresponding diagrams in Fig. \ref{fig:2D_spectra}(c). 
The spectral peak amplitudes in this regime align well with analytical solutions for FWM under $\delta$-function pulse excitation.
However, as discussed earlier, we need to use a reduced value of $\mu_{02}$ to compensate for the lower spectral intensity at the detuned transition.

\begin{figure}
\includegraphics{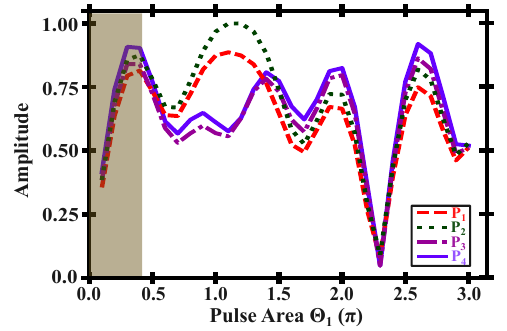}
\caption{Variation in the maximum amplitudes of each peak within the 2D spectra in response to changes in the pulse area of the first pulse $\Theta_{1}$, while the subsequent pulses remain within the perturbative regime. The shaded region indicates the transition from perturbative predictions to the onset of saturation.}
\label{fig:Pulse_Area_effect}
\end{figure}

We plot the maximum amplitudes of each peak as a function of $\Theta_{1}$ in Fig. \ref{fig:Pulse_Area_effect}.
The shaded region in the plot highlights the transition from a perturbative to saturation behavior, which was previously observed in both simulation and experiment \cite{Tripathi2025}.
At higher pulse areas, the system exhibits nonmonotonic behavior.
The origin of this behavior can be understood by examining the single-pulse excitation results shown in Fig. \ref{fig:Norm_pop}(d).
DSFDs clearly indicate that peaks $P_1$ and $P_2$ originate from pathways involving the coherence $\rho_{01}$ formed after the first pulse, whereas $P_3$ and $P_4$ are associated with $\rho_{02}$.  
Accordingly, the amplitudes of the peaks in the 2D spectra reflect the magnitude of their respective coherences after excitation by the first pulse.
Our simulation results confirm this correspondence: the amplitudes of peaks $P_1$ and $P_2$ closely follow the pattern of $\abs{\rho_{01}}$, while $P_3$ and $P_4$ follow the pattern of $\abs{\rho_{02}}$ observed in the single pulse excitation case. 
Although the DSFDs are only applicable in the perturbative regime, the correspondence between the single-pulse and 2DCS calculations highlights the usefulness of the qualitative interpretation of 2D spectra based on the DSFDs even in the nonperturbative regime.

These findings highlight the potential to selectively tune the relative amplitudes of spectral peaks, i.e., control specific coherent pathways, by varying the pulse area of the first excitation pulse within a range of $0.5 \pi$ to $2.1 \pi$. 
While modifying the intensity of the first pulse enables switching the more prominent peak between $P_1$ ($P_2$) and $P_4$ ($P_3$), it cannot amplify and suppress individual peaks, with respect to others.
We explore this possibility by systematically varying the pulse areas of the first three pulses over the same range ($0.5\pi$ to $2.1\pi$).
We take the brute-force approach to calculate the optimal pulse areas for isolating individual peaks in the absence of an intuitive, analytical method of prediction.
We fix $\Theta_4 = 0.1\pi$, which is within the perturbative limit.
Thus, we establish a methodological bridge between the collinear and noncollinear geometries in the context of 2DCS experiments \cite{Tripathi2025}. 
An additional advantage is that there is no need to vary $\Theta_4$ in our simulations, thereby reducing simulation time.

\section{Coherent Control Results}
\label{Results}
\subsection{Intensity Control}
\label{I_control}

\begin{figure}
\includegraphics{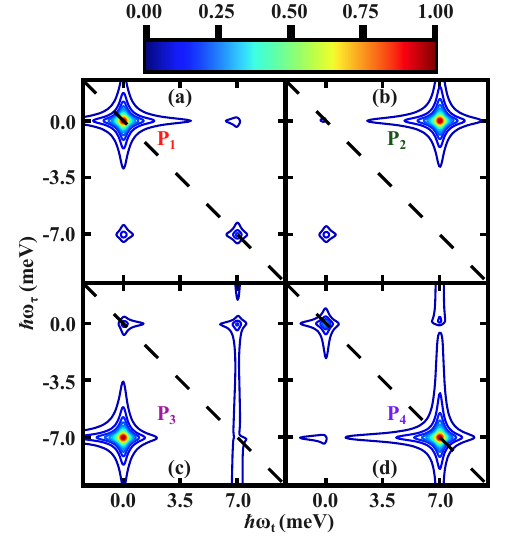}
\caption{Simulated rephasing 2D spectra demonstrating Coherent control in V-Level system: variations (a) to (d) with different pulse area combinations. The spectra are individually normalized. Color map indicates the amplitude.}
\label{fig:CC_2D}
\end{figure}

The ability to selectively enhance or suppress individual peaks in the rephasing 2D spectra is demonstrated in Fig. \ref{fig:CC_2D}.
The pulse area values to obtain these spectra are listed in Table \ref{tab:Pulse_areas}.

\begin{table}[b]
\caption{\label{tab:table2}Pulse area combination to control different peaks' intensity of V-level system's 2D spectra.}
\begin{ruledtabular}
\begin{tabular}{cccccc}
 Figure&$\Theta_{1}(\pi)$&$\Theta_{2}(\pi)$&$\Theta_{3}(\pi)$&
 Peak visibility(\%) \\
\hline
\ref{fig:CC_2D}(a) & 1.1&1.3&1.6& 95& \\
\ref{fig:CC_2D}(b) &1.1&0.8&1.5& 98 \\
\ref{fig:CC_2D}(c)&1.5&1.5&0.9& 95 \\
\ref{fig:CC_2D}(d)& 2.1&0.9&0.9& 91
\end{tabular}
\end{ruledtabular}
\label{tab:Pulse_areas}
\end{table}

Figure \ref{fig:CC_2D}(a) highlights $P_{1}$ as the only prominent peak, illustrating a scenario where the excitation pulses result in a constructive interference between pathways leading to the peak $P_{1}$, which corresponds to the first excited state, and nearly destructive interference of pathways leading to all other peaks ($P_{2}$, $P_{3}$, and $P_{4}$).
We can state that, from the eight transition pathways illustrated in Fig. \ref{fig:2D_spectra}(c), only the first two pathways (labeled as $P_{1}$) are isolated.
This observation confirms that, in this specific scenario, the nonlinear signal arises primarily from the coherence $\rho_{01}$ formed after interactions with the first and third pulses.
Similarly, Figs. \ref{fig:CC_2D}(b)-\ref{fig:CC_2D}(d) each show the selective enhancement of a single peak - $P_2$, $P_3$, and $P_4$, respectively, results from the isolation of their corresponding pathways in Fig. \ref{fig:2D_spectra}(c).
These results demonstrate a high degree of coherent control, where the excitation conditions can be tuned to favor specific quantum pathways, enabling one peak to dominate while suppressing the others.

We quantify the degree of peak selectivity in 2D spectra through peak visibility, which we define as the ratio of a peak's intensity to the sum of the intensities of all peaks in a 2D spectrum, expressed as a percentage:
\begin{equation}
    PV_{P_{i}} = \frac{I_{P_{i}}}{\sum_{i=1}^{4} I_{P_{i}}} \times 100\%.
\end{equation}
In the equation, $I_{P_{i}}$ represents the highest intensity of peak $P_{i}$ in a 2D spectrum. 
The highest intensity of each peak is determined by the square of its maximum amplitude.
The peak visibility of the dominant peaks for each spectrum in Fig. \ref{fig:CC_2D} are listed in Table \ref{tab:Pulse_areas}.
We achieve a peak visibility of more than $90\%$ for each of the dominants peaks as shown in the spectra in Fig. \ref{fig:CC_2D}.
This observation indicates a remarkable degree of control over the coherence dynamics in the 2D spectra of the V-level system.

\begin{figure}
\includegraphics{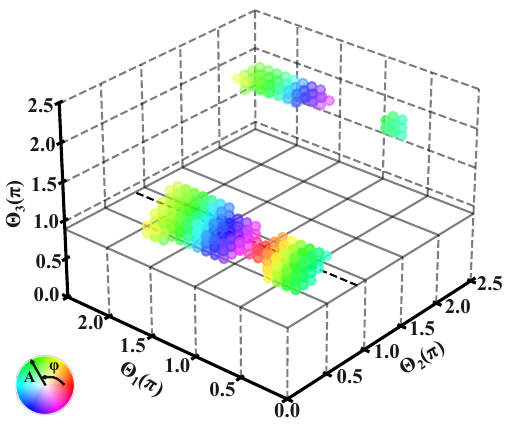}
\caption{3D hue-saturation plot of $P_{4}$ diagonal peak intensity: hue represents the phase and saturation indicates the intensity.
Specifically, we have selected points with $I_{P_4} \geq 0.7$ times its maximum value across all spectra. 
The plot demonstrates phase variation with $\Theta_{1}$.}
\label{fig:Phase_control}
\end{figure}

We note that it is possible to modify the relative strength of the peaks by varying $T$.
Specifically, in the perturbative regime, the cross-peak amplitudes oscillate with a period of $T=1/\Delta$ while having a negligible effect on diagonal peaks \cite{Nardin2014}.
However, as we discuss in Appendix \ref{app:Role_T}, we find that the above does not hold true for nonperturbative excitation.
Nevertheless, varying $T$ does not seem to affect the peak visibility discussed here; a new set of excitation pulse areas can effectively isolate individual peaks for any $T$ significantly shorter than relevant population decay times.


\subsection{Phase Control}
\label{Phase_control}
We can also precisely control the signal phase in addition to the intensity variations discussed in Sec. \ref{I_control}.
This capability is illustrated in Fig. \ref{fig:Phase_control}, where a 3D plot depicts the peak intensity of diagonal peak $P_{4}$ in relation to the pulse area of the initial three pulses.
This plot demonstrates that a significant number of points cluster around a pulse area of $0.9\pi$ for both the second and third pulses. 
Intriguingly, the plot reveals that, while the intensity of diagonal peak $P_{4}$ remains relatively constant, its phase undergoes a complete cycle; the phase varies over the entire range of $0$ to $2\pi$ as $\Theta_1$ is varied. 

\begin{figure}
\includegraphics{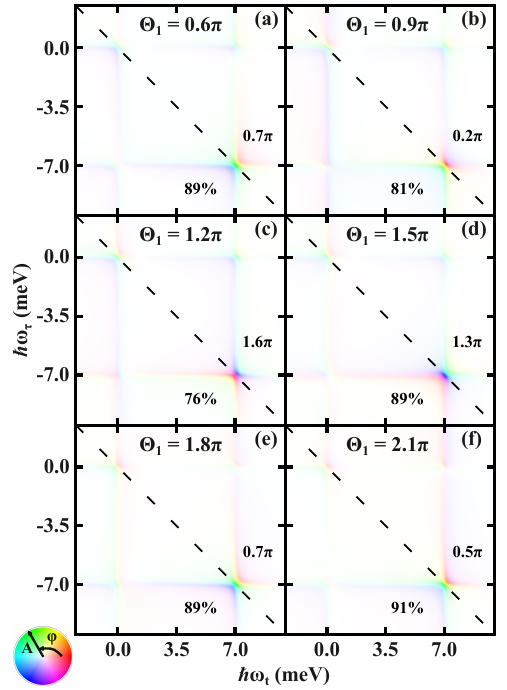}
\caption{2D spectra with amplitude normalized to the global maximum value across these spectra, with hue and saturation indicating phase and amplitude, respectively. The phase of the most intense point and peak visibility of peak $P_4$ are shown for each spectrum.
The specific values of $\Theta_1$ are indicated in each subfigure; $\Theta_2=\Theta_3=0.9\pi$ for all the spectra.}
\label{fig:Phase_2D}
\end{figure}

In Fig. \ref{fig:Phase_control}, a distinct line parallel to $\Theta_{1}$ is observed at $\Theta_{2}$ = $\Theta_{3}$ = $0.9\pi$, along which six points were selected to highlight the phase evolution of diagonal peak $P_4$. 
We plot these 2D spectra in Fig. \ref{fig:Phase_2D}. Each spectrum is normalized to the global maxima across these spectra. 
As shown in Fig. \ref{fig:Phase_2D}, each successive spectrum exhibits a distinct phase of the diagonal peak $P_{4}$, while the intensity of the peak does not change significantly. 
Similarly, the phase modulation of the other peaks can also be achieved by adjusting the pulse area combination of the initial three pulses, while keeping the intensity of the corresponding peaks nearly constant (data not shown).
We note that the phase dependence for a peak on the intensity of the excitation pulses can be significantly more complicated compared to the roughly linear dependence on the intensity of a single pulse, as shown in Fig. \ref{fig:Phase_control} for peak $P_4$.


\section{Conclusions}
\label{Conclusion}
In this study, we have demonstrated a straightforward and widely applicable approach to control the coherent evolution of multilevel systems. 
The state of a V-type three-level system after excitation by a single, intense excitation pulse was found to be significantly affected by the product of the pulse duration and the energy separation between the excited states.
This insight motivated the 2DCS study in the high-intensity regime.
We were able to turn the contribution of specific quantum pathways on or off through pulse area combinations of the excitation pulses, which is demonstrated through a control over intensity of individual peaks. 
Furthermore, we demonstrated that the phase of individual peaks can be manipulated without affecting their amplitude significantly. 
These results showcase a new strategy to control the coherent optical response in V-level systems.


A particularly interesting application of this work would be in optical control of the radiated FWM signal.
The various phase-matching directions in a FWM experiment performed in the noncollinear geometry correspond to different types of signals.
For example, the rephasing signal discussed in this work is emitted in the direction $-\vb{k_A}+\vb{k_B}+\vb{k_C}$, where the wavevectors indicate the propagation direction of the three excitation pulses.
By changing the pulse areas of the excitation pulses, the frequency and phase of the radiated FWM signal can be controlled.
The ability to control coherent dynamics through intensity-dependent nonlinear interactions may also provide a potential route toward photon locking, where nonlinear coupling stabilizes specific photon populations or phases \cite{Bayer2009}.

It is also instructive to contrast this work with the previously reported coherent-control approaches.
Polarization-based protocols can tune or suppress specific transitions in systems with orthogonal transition dipoles \cite{Wang2005}.
Even in the absence of orthogonal transition dipoles, polarization scheme with specific sequences in pump-probe geometry can suppress diagonal peaks, thereby revealing weaker cross peaks \cite{Farrell2022}.
Chirped-pulse and other complex pulse-shaping techniques can selectively modify transition strengths and thereby modulate 2D spectral features \cite{Lim2011,Tollerud2014,Marroux2016}.
Still, isolating a single diagonal or cross-diagonal feature has not been possible.
In contrast, our approach is straightforward to implement, works in the high-intensity regime where signal amplitudes and, consequently, SNRs are significantly higher, and enables us to isolate individual spectral features. 
Such strong-field control may also provide opportunities to reveal and manipulate vibrational dynamics that remain inaccessible in the weak-excitation regime even with relatively longer pulses \cite{Chen2017}.
Furthermore, we demonstrate control of the signal phase of the individual peak, which had not been shown previously.

The proposed approach is expected to be relevant for a wide class of systems exhibiting V-type level structures, including semiconductor quantum dots \cite{Stievater2002}, transition metal dichalcogenides (TMDs) \cite{Hao2016a}, quantum well nanostructures, and atomic and molecular systems \cite{Meshulach1998,Marroux2016,Lim2011}.
In such systems, factors such as inhomogeneous broadening \cite{Suzuki2018} and coupling to additional states (e.g., biexciton levels \cite{Suzuki2016}) can lead to more complex dynamics.
While these effects are not explicitly considered here, they provide a natural extension of the present framework and offer opportunities to explore pathway-selective control in realistic systems.
However, as a consequence of the high-intensity excitation, to accurately predict the system's response, one might have to incorporate many-body effects that dominate in semiconductor nanostructures \cite{Chemla2001} and other strong-field effects such as multiphoton absorption \cite{Abella1962}, finite bleaching \cite{Kong2007}, and exciton-exciton annihilation \cite{Sun2014}, which have not been considered here.

\begin{acknowledgments}
The authors acknowledge support from the Science and Engineering Research Board (SERB), New Delhi under Project No. CRG/2023/003263.
K. K. M. acknowledges the Ministry of Education, Government of India for support from the Prime Minister’s Research Fellows (PMRF) Scheme.
\end{acknowledgments}

\appendix
\section{OBEs for V-level system}
\label{app:OBEs_V}
The OBEs, initially formulated to describe the dynamics of a two-level systems interacting with electromagnetic fields, can be extended to multiple level systems. By adding new elements to account for electromagnetic couplings and relaxation processes between energy levels, extended OBEs for V-level systems is reduced to a set of equations shown below.
\begin{subequations}
    \begin{align}
       {\dot{\rho}}_{01} &= -(i \omega_{01} + \gamma_{01})\rho_{01} - \frac{i}{\hbar} V_{01}(\rho_{11}- \rho_{00}) \\
       &\quad- \frac{i}{\hbar} V_{02}\rho_{21} \nonumber, \\ 
       {\dot{\rho}}_{02} &= -(i \omega_{02} + \gamma_{02})\rho_{02} - \frac{i}{\hbar} V_{02}(\rho_{22}- \rho_{00}) \\
        & \quad - \frac{i}{\hbar} V_{01}\rho_{12}\nonumber, \\
        {\dot{\rho}}_{11} &= -(\gamma_{1})\rho_{11} - \frac{i}{\hbar} (V_{10}\rho_{01}- \rho_{10}V_{01}), \\
        {\dot{\rho}}_{22} &= -(\gamma_{2})\rho_{22} - \frac{i}{\hbar} (V_{20}\rho_{02}- \rho_{20}V_{02}), \\
          {\dot{\rho}}_{12} &= -(i \omega_{12} + \gamma_{12})\rho_{12} - \frac{i}{\hbar} (V_{10}\rho_{02}- V_{02}\rho_{10}),\\
          {\dot{\rho}}_{00} &= (\gamma_{1})\rho_{11} + (\gamma_{2})\rho_{22} + \frac{i}{\hbar} (V_{10}\rho_{01}- \rho_{10}V_{01}) \\
  &\quad + \frac{i}{\hbar} (V_{20}\rho_{02}- \rho_{20}V_{02}) \nonumber.
    \end{align}
\end{subequations}

In the set of OBEs, the terms $\rho_{jk}$ represent the elements of the density matrix. For cases where $j=k$, these terms denote the populations of the energy state $\ket{j}$. Conversely, when $j\neq k$, they describe the coherences between energy states $\ket{j}$ and $\ket{k}$. The interaction between the system and the electromagnetic field is captured by $V_{jk}= -\mu_{jk}\cdot E(t)$. Here, $E(t)$ symbolizes the electric field, defined as $E(t)=\frac{1}{2}G(t)(\exp{i\omega_{L}t} + c.c.)$, where $G(t)$ signifies the pulse envelope, and $\omega_{L}$ represents the central frequency of the light pulse. Additionally, $\gamma_{jk}$ denotes the decay rate of the coherence term $\rho_{jk}$, and $\omega_{jk}$ is defined as the transition frequency between the energy states $\ket{j}$ and $\ket{k}$. 

\begin{figure}[t]
\includegraphics{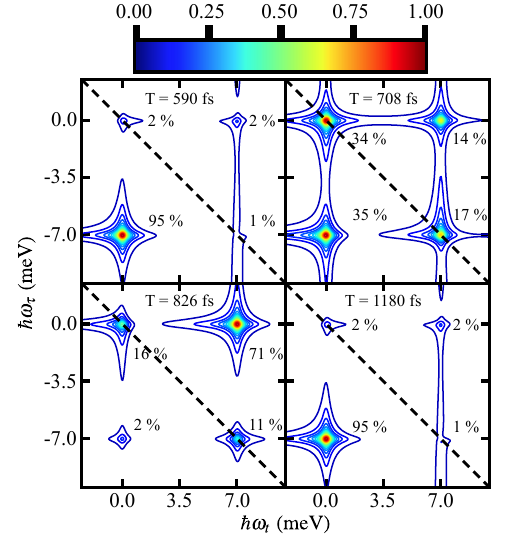}
\caption{Effect of varying $T$ on the rephasing 2D spectra for the pulse area combination used in Fig. \ref{fig:CC_2D}(c)}
\label{fig:Role_T}
\end{figure}

\section{Role of the Inter-Pulse Delay $T$ in Coherent Control Response}
\label{app:Role_T}
Varying delay $T$ primarily affects the off-diagonal peaks in the perturbative regime; the cross peaks oscillate with a period that is inverse of the detuning between the excitation states \cite{Nardin2014}.
In order to understand the effect of varying $T$, we simulated 2D spectrum shown in Fig. \ref{fig:CC_2D}(c) for a range of values between 590-1770 fs, which spans one to three oscillation periods associated with the 7 meV energy separation between the two excited states.
We plot a subset of this data in Fig. \ref{fig:Role_T} to highlight the effect of varying the inter-pulse delay $T$, while maintaining the pulse-area combination as defined in Table \ref{tab:Pulse_areas} for Fig. \ref{fig:CC_2D}(c). 
The amplitudes and visibilities of all peaks evolve with a periodicity of $T=1/\Delta=590$ fs. 
Thus, we conclude that for high-intensity excitation, although the periodicity is maintained, varying $T$ affects all the peaks significantly.

Consequently, the value of $T$ should not significantly affect the coherent control phenomena discussed in the main text as long as $T$ is significantly less than the relevant population decay timescale.
One would, however have to identify a different set of pulse-intensities to achieve a desired result.

\bibliography{Coherent_control_PRA}

\end{document}